\documentclass[a4paper,12pt,english]{article}
\usepackage{amssymb}
\usepackage{amsfonts}
\usepackage{graphicx}
\usepackage{amsmath}
\usepackage{color}
\usepackage{hyperref}

\newcommand{\be}{\begin{equation}}
\newcommand{\ee}{\end{equation}}
\newcommand{\badat}{\begin{alignedat}}
\newcommand{\eadat}{\end{alignedat}}
\newcommand{\virg}{\hspace{1 mm}, \hspace{8 mm}}
\newcommand{\rd}{\mathcal}

\begin{document}

%\tableofcontents

\begin{titlepage}
\begin{center}

\noindent{{\LARGE{WAdS$_3$/CFT$_2$ correspondence in presence of bulk massive gravitons}}}

\smallskip
\smallskip

\smallskip
\smallskip
\smallskip
\smallskip
\noindent{\large{Laura Donnay$^{1}$, Gaston Giribet$^{1,2}$}}

\smallskip
\smallskip

\end{center}

\smallskip
\smallskip
\centerline{$^1$ Universit\'{e} Libre de Bruxelles and International Solvay Institutes}
\centerline{{\it ULB-Campus Plaine CPO231, B-1050 Brussels, Belgium.}}

\smallskip
\smallskip
\centerline{$^2$ Departamento de F\'{\i}sica, Universidad de Buenos Aires and IFIBA-CONICET}
\centerline{{\it Ciudad Universitaria, Pabell\'on 1, 1428, Buenos Aires, Argentina.}}

\bigskip

\bigskip

\bigskip

\bigskip

\begin{abstract}

In a previous paper, it has been shown that the entropy of non-extremal
black holes in Warped Anti-de Sitter (WAdS) spaces in massive gravity can
be computed microscopically in terms of a dual conformal field theory.
Here, we extend this computation to a set of asymptotic boundary
conditions that, while still gathering the WAdS$_3$ black holes, also
allow for new solutions that are not locally equivalent to WAdS$_3$ space,
and therefore are associated to the local degrees of freedom of the theory
(bulk massive gravitons). After presenting explicit examples of such
geometries, we compute the asymptotic charge algebra and show that it is
generated by the semi-direct sum of Virasoro algebra and an affine
Kac-Moody algebra. The value of the central charge turns out to be exactly
the one that leads to reproduce the entropy of the WAdS$_3$ black holes.
This result probes the WAdS$_3$/CFT$_2$ correspondence in presence of bulk
gravitons.

\end{abstract}

\end{titlepage}
%%%%%%%%%%%%%%%%%%%%%%%%%%%%%%%%%%%%%%%%%%%%%%%%%%%%%%%%%%%%%%%%%%%%%%%%

%\newpage

\section{Introduction}

Warped Anti-de Sitter spaces (WAdS) are squashed or stretched deformations of $3$-dimensional AdS space that preserve the $SL(2,\mathbb{R})\times U(1)$ subgroup 
of the AdS$_3$ isometries. These spaces happen to be solutions of a large class of theories, including string theory, and appear in several contexts, such as in 
the  Kerr/CFT correspondence \cite{application1}. 

In Ref. \cite{Aninnos}, a WAdS/CFT correspondence was conjectured, stating that quantum gravity in asymptotically WAdS spaces is dual to a $2$-dimensional 
conformal field theory (CFT).  This represents one of the most interesting attempts to extend AdS/CFT correspondence to non-AdS, less symmetric spaces. One of the 
reasons why WAdS/CFT is interesting is the existence of black hole solutions \cite{BHs} that are locally equivalent\cite{Aninnos} to WAdS$_3$ and are also 
asymptotically WAdS$_3$. Therefore, the WAdS/CFT correspondence provides a setup to address the holographic description of non-extremal, non-AdS black holes. In 
a previous paper\cite{nos2}, we showed that, for the particular case of massive gravity, the WAdS$_3$ black hole entropy could be microscopically described in 
terms of a dual CFT$_2$. More precisely, we showed that the states of the dual theory that corresponds to the black hole configurations organize in highest 
weight representations of two mutually commuting Virasoro algebras, and that the Cardy formula for the asymptotic growth of such states exactly reproduces the 
black hole entropy. 

A crucial ingredient in the analysis done in Ref. \cite{nos2} is the asymptotic boundary conditions considered. The question we are concerned with in this 
paper is 
whether a similar WAdS/CFT computation can still be carried out for a set of boundary conditions which, on top of the black hole solutions, admit also the presence of bulk gravitons\footnote{The term {\it bulk graviton} is used in contraposition to the term {\it boundary graviton}, which is usually employed to refer to degrees of freedom associated to the representations of the asymptotic isometries; the latter being also present in theories with no local degrees of freedom.}. That is, we will consider a one-parameter family of boundary conditions which, while accommodating the WAdS$_3$ black holes configurations, also gather non-locally WAdS$_3$ solutions. The result we will obtain strongly suggests that WAdS/CFT correspondence is still valid in presence of bulk gravitons.

\section{WAdS$_3$ black holes and massive gravity}

We are interested in the black hole geometries \cite{BHs, Clement}, 
\be
\badat{2}
\frac{ds^2}{l^2}=  dt^2 + \frac{dr^2}{(\nu^2+3)(r-r_+)(r-r_-)}+\left( 2\nu r -\sqrt{r_+r_-(\nu^2+3)}\right) dt d\varphi\\
+\frac{r }{4}\left( 3(\nu^2-1)r+(\nu^2+3)(r_++r_-)-4\nu \sqrt{r_+r_-(\nu^2+3)}\right) d\varphi^2,
\label{warped}
\eadat
\ee
which at large distance asymptote WAdS$_3$ space. Metric (\ref{warped}) represents black holes with inner and outer horizons located at $r_+$ and $r_-$, 
respectively. $\nu $ is the parameter that controls the stretching deformation, with AdS$_3$ corresponding to $\nu =1$; $l $ is the curvature radius. 

Black holes (\ref{warped}) can be constructed by means of global identifications from the WAdS$_3$ space \cite{Aninnos}. This orbifold type construction of these 
geometries yields two geometrical temperatures, given by the inverse of the identification periods; namely
\begin{equation}
T_{\text{L}} = \frac{\nu^2+3}{8\pi l^2} (r_+ + r_--\frac{1}{\nu } \sqrt{(\nu^2+3)r_+r_-}), \ \ \ T_{\text{R}} = \frac{\nu^2+3}{8\pi l^2} (r_+ - 
r_-).\label{geometrica}
\end{equation} 

As mentioned before, WAdS$_3$ spaces and black holes appear in a large class of theories, such as topologically massive gauge theories, string theory, and many others. A minimal set up in which these geometries appear as solutions is $3$-dimensional pure gravity, provided one gives to the graviton a mass
\begin{equation}
m^2 = \frac{(20\nu^2-3)}{2l^2}, 
\end{equation}
and the cosmological constant takes the value
\begin{equation}
\Lambda = \frac{(4\nu^4-48\nu^2+9)}{2l^2(20\nu^2-3)}.
\end{equation}

In three dimensions, there is a consistent manner to give mass to the graviton preserving parity invariance. This amounts to consider the higher-order action
\begin{equation}
{\mathcal I} = \frac{1}{16\pi G} \int d^3 x \sqrt{-g} \left( R -2\Lambda - \frac{1}{m^2} (R_{\mu \nu }R_{\mu \nu }-\frac{3}{8} R^2)\right) , \label{I}
\end{equation}
which is a fully covariant extension of the spin-$2$ massive Fierz-Pauli theory \cite{NMG}.

The mass ($M$) and angular momentum ($J$) of black holes (\ref{warped}) in massive gravity (\ref{I}) can be computed by different methods. Their expressions in 
terms of $r_-$ and $r_+$ can be found, for instance, in Ref. \cite{nos2}. On the other hand, the entropy of these black holes can be computed with the Wald 
formula, and in terms of the mass and the angular momentum, it takes the form
\begin{equation}
S_{\text{BH}} = \frac{4\pi l\nu }{(\nu^2 + 3)} (M +\sqrt{ M^2 -k J}) , \label{V2}
\end{equation}
with $k = {4 \nu (3+\nu^2)}/{(G l(20 \nu^2 -3))}$.

%We will show later how the black hole entropy (\ref{V2}) can be recovered 
%in terms of the two temperature (\ref{geometrica}) in the dual theory.

\section{Asymptotic symmetries and charges}

Black hole solutions (\ref{warped}) break the $SL(2,\mathbb{R})\times U(1)$ isometry group down to $U(1)\times U(1)$. The latter group is generated by the two 
Killing vectors $\xi^{(1)} = \partial_t $, $\xi^{(2)} = - \partial_{\varphi}$. We will see, however, that the asymptotic isometry group is infinite-dimensional. 
The asymptotic boundary conditions we consider here are defined as follows: We first consider as reference background the metric (\ref{warped}) with $r_-=r_+=0$; 
we denote the metric of such geometry by $ds_0^2$. Then, the set of geometries to be considered are defined as those deformations of the form $ds^2=ds^2_0+\delta 
g_{\mu \nu }dx^{\nu }dx^{\mu }$ that respect the following fall-off conditions
\begin{eqnarray}
\begin{alignedat}{2}
&\delta g_{tt}={\mathcal O}(r^{ -3}) \virg \delta g_{tr}={\mathcal O}(r^{\delta -4}) \virg \delta g_{t\varphi 
}={\mathcal O}(r^{\delta - 1}),  \\
&\delta g_{rr}={\mathcal O}(r^{\delta -4}) \virg \delta g_{r\varphi}={\mathcal O}(r^{\delta -2}) \virg \delta 
g_{\varphi \varphi}={\mathcal O}(r^{\delta }).
\label{falloff2}
\end{alignedat}
\end{eqnarray}
where $\delta $ is a real parameter that here we will assumed greater that $1$. For $\delta >1 $, asymptotic conditions (\ref{falloff2}) exhibit components that 
fall off slower that those of \cite{Compere3, nos2}.

Boundary conditions (\ref{falloff2}) are similar to those considered in the literature \cite{Enoc} for the parity-odd theory, and it is worthwhile comparing such 
definition with the asymptotic boundary conditions of Refs. \cite{Compere3, nos2}. Fall-off conditions (\ref{falloff2}) accommodate, in particular, the black 
hole solutions (\ref{warped}), but also include other solutions: An example of such a solution is given by the Kerr-Schild ansatz \begin{equation}
ds^2= ds^2_0 + e^{\omega t} r^{-\frac{2\nu\omega }{3+\nu^2}} h(u)\  k_{\mu}k_{\nu} dx^{\mu} dx^{\mu},
\label{warpeddeformed}
\end{equation}
with the null vector $k_{\mu }$
\begin{equation}
k_{\mu} dx^{\mu }= \frac{2}{3+\nu^2} \frac{dr}{r^2}- d\varphi \ , \label{warpeddeformed2}
\end{equation}
$h(u)$ being an arbitrary periodic function of the variable $u\equiv \varphi + 2/((\nu^2+3)r) $ (hereafter, we set $l=1$) and $\omega $ being a solution of the 
polynomial equation
\begin{equation}
P(\nu , \omega )\equiv -4\nu\omega^2-6\omega+10\omega\nu^2+16\nu^3-\omega^3=0. \label{polinomio}
\end{equation}

Solution (\ref{warpeddeformed})-(\ref{polinomio}) obeys the boundary conditions (\ref{falloff2}) with $\delta = -2\nu \omega /(\nu^2+3)$ and are not locally equivalent to WAdS$_3$.

Asymptotic Killing vectors respecting the set of boundary conditions (\ref{falloff2}) are given by 
\begin{equation}
\begin{alignedat}{2}
&\ell_n =\frac{4\nu n^2}{(3+\nu^2)}\frac{1}{r}e^{-in\varphi} \partial_t   +\frac{2\nu 
n^2-r^2(3+\nu^2)}{(3+\nu^2)}\frac{1}{r^2}e^{-in\varphi}\partial_{\varphi } -i n re^{-in\varphi }\partial_r + ... \label{Kil} \\ 
&t_n = ( e^{in\varphi}+\rd O(r^{-1})) \partial_t + ... 
\end{alignedat}
\end{equation}
where $n \in \mathbb Z$, and where the ellipses stand for sub-leading terms. In particular, these asymptotic Killing vectors include the exact Killing vectors 
$\ell _0 = -\partial_\varphi$ and $t_0 =\partial_t$. The full set of asymptotic Killing vectors satisfy the algebra
\begin{equation}
i[\ell_m,\ell_n]=(m-n)\ell_{m+n}, \ \ \ i[\ell_m,t_n]=-n t_{m+n}, \ \ \ i[t_m,t_n]=0 , \label{Wito} 
\end{equation}
which is a semi-direct sum of Witt algebra and the loop algebra of $u(1)$. 

The next step is to compute the charges associated to vectors $\ell_n $ and $t_n$. This computation can be performed as explained in Ref. \cite{nos2}; that is, 
by 
resorting to the Brandt-Barnich-Comp\`{e}re covariant formalism \cite{Barnich:2001jy,Barnich:2007bf}, using the metric $ds^2_0$ as reference background, and the 
expression of the charges derived in Ref. \cite{NPY} for massive gravity. If we denote by $L_n$ and $P_n$ the charges associated to ${\ell_n}$ and ${t_n}$ 
respectively, then the algebra of charges reads 
\begin{eqnarray}
\begin{alignedat}{2}
i\{L_m,L_n\}=(m-n)L_{m+n}+\frac{c}{12} m^3 \delta_{m+n,0} ,  \\
i\{L_m,P_n\}=-n P_{m+n},  \ \  i\{P_m,P_n\}=  \frac{k}{2} m \delta_{m+n,0}, 
\label{74}
\end{alignedat}
\end{eqnarray}
where
\begin{equation}
c=\frac{96 \nu^3}{G (20 \nu^4 +57 \nu^2  -9)} ,
\end{equation}
and where $k$ coincides with the value already appearing in (\ref{V2}). This algebra is a semi-direct sum of Virasoro algebra with central charge $c$ and the affine $\hat{u}(1)_k$ Kac-Moody algebra of level $k$. 

In order to organize the states that correspond to the black hole configurations, we define the Sugawara operators \cite{blago, nos2} 
\begin{equation}
L^{-}_n \equiv  \frac{1}{k} \sum_{m\in\mathbb{Z}} : P_{-n-m} P_{m} :  \ , \ \ \ \ L_{n}^{+} \equiv L^{-}_{n} + L_{n} . \label{TY1}
\end{equation}
These operators both obey the Virasoro algebra with central charges $c^-=1$ and $c^+=c-1$, respectively; they also satisfy $
[L^{+}_n,L^{-}_m] = 0 $. One can also verify that these operators are bounded from below \cite{nos2} when evaluated on the black hole spectrum (see Fig. 1); 
namely,
\begin{equation}
L_0^{+} = \frac{1}{k} M ^2 - J \geq 0 , \ \ \ \ L_0^{-} = \frac{1}{k} M ^2 \geq 0.
\end{equation}

\begin{figure}
\includegraphics[width=7in]{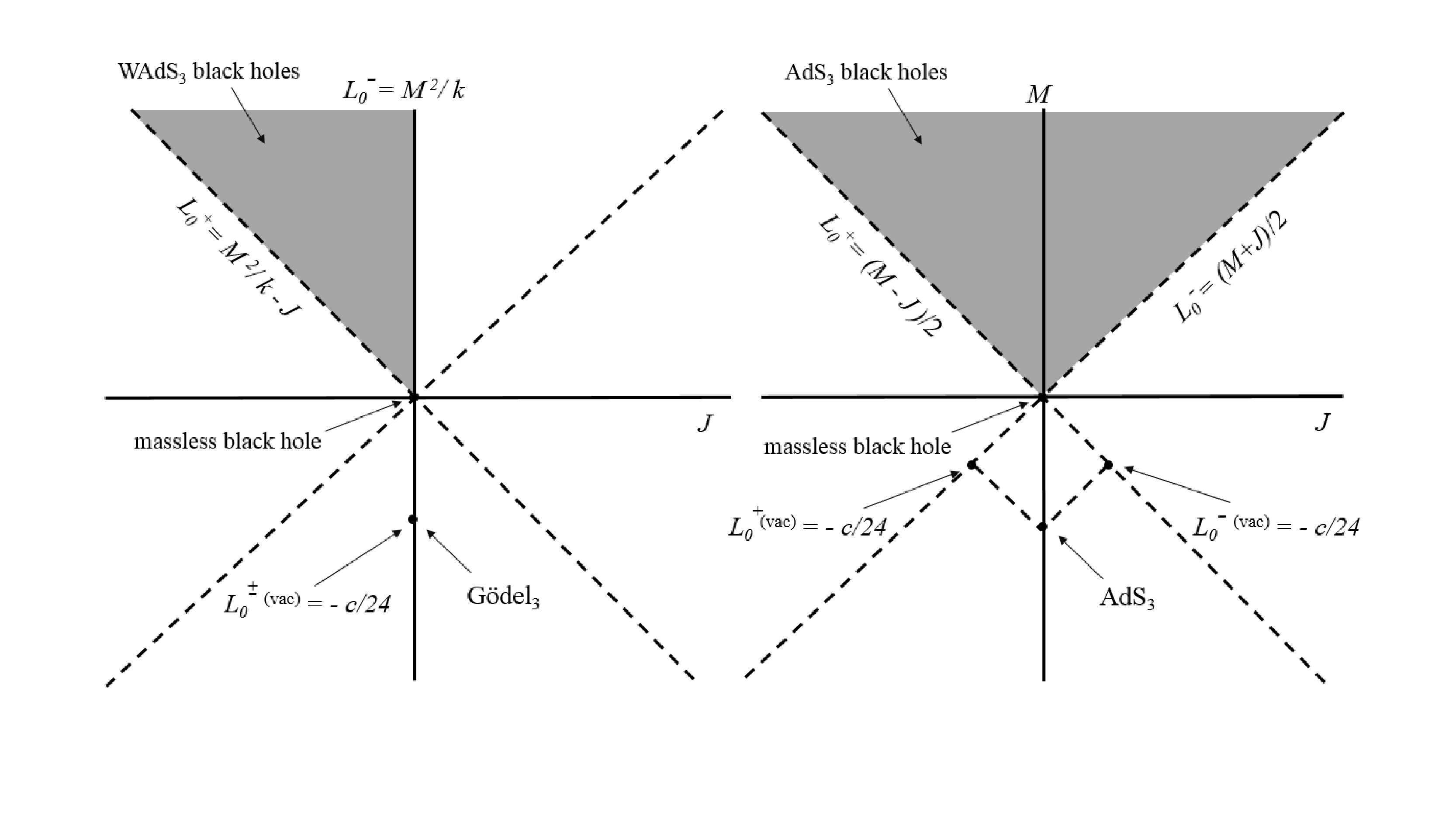}
\caption{The spectrum of WAdS$_3$ black holes is qualitatively similar to that of the AdS$_3$ black holes. The background geometry with 
respect to which one computes the conserved charges and the asymptotic boundary conditions is the configuration with $r_+=r_-=0$. On the other hand, the spectrum exhibits a gap with the ground state being the configuration with $L_0^{\pm }=-c/24$; in the case of AdS$_3$ black holes this corresponds to AdS$_3$ space, while for the 
WAdS$_3$ black holes this corresponds to G\"odel space.}
\label{aba:fig1}
\end{figure}

\section{Microscopic black hole entropy}

The existence of two mutually commuting Virasoro algebras whose highest weight representations are consistent with the black hole spectrum invites to check whether the Cardy entropy formula reproduces the black hole entropy (\ref{V2}). Then, we consider
\begin{equation}
S_{\text{CFT}} = 2\pi \sqrt{\frac{c_{\text{eff}}^{-}}{6} {L}_0^{-} } + 2\pi \sqrt{\frac{c_{\text{eff}}^{+}}{6} {L}_0^{+} }, \label{jk}
\end{equation}
where ${c}_{\text{eff}}^{\pm }\equiv -24{L}_0^{\pm (\text{vac})}$ are the effective central charges, ${L}_0^{\pm (\text{vac})}$ being the minimum values that 
charges ${L}_0^{\pm}$ take in the spectrum. That is, the values ${L}_0^{\pm (\text{vac})}$ are the charges evaluated on the geometry that is to be identified as 
the ground state of the black hole spectrum (i.e. the vacuum). Such geometry has to exhibit the full $SL(2,\mathbb{R})\times U(1)$ isometry group, and can be 
argued to be the timelike WAdS$_3$ space, which corresponds to the 3-dimensional section of G\"odel spacetime \cite{DHH, nos2}. The latter is precisely the 
geometry that reduces to global AdS$_3$ when $\nu =1$, and its geometry corresponds to \cite{Donnayt} 
\begin{equation}
{L}_0^{\pm (\text{vac})}=-\frac{c}{24}. \label{cp}
\end{equation}

This implies that the effective central charges associated to the Virasoro algebras generated by operators $L_n^{\pm}$ are given by $c_{\text{eff}}^{\pm } =c$. This explains the fact that black hole entropy (\ref{V2}) can be expressed in terms of the geometric temperatures (\ref{geometrica}) by
\begin{equation}
S_{\text{BH}} = \frac{\pi^2 }{3} ( c_{\text{eff}}^{- } \ T_{\text{L}} + c_{\text{eff}}^{+ } \ T_{\text{R}} ) = \frac{\pi^2 }{3} c (T_{\text{L}} + T_{\text{R}}) ,
\end{equation}
with the value of $c$ conjectured in Ref. \cite{Goya}.

With (\ref{cp}), one finds that Cardy formula (\ref{jk}) exactly matches the black hole entropy (\ref{V2}); namely one finds $
S_{\text{CFT}} = S_{\text{BH}}$. This result generalizes our previous result of \cite{nos2} to the one-parameter family of boundary conditions (\ref{falloff2}), 
and can be regarded as a consistency check of WAdS/CFT correspondence in the case of boundary conditions that accommodate bulk gravitons.

\section*{Acknowledgements}

L.D.~is a research fellow of FRIA Belgium and her work is supported in part by IISN-Belgium and by
``Communaut\'e fran\c caise de Belgique - Actions de Recherche
Concert\'ee.

\providecommand{\href}[2]{#2}\begingroup\raggedright\endgroup

\end{document}